\documentstyle[prl,aps,multicol,amssymb,amsmath,epsf,psfrag]{revtex}

\def\unity{\mbox{\small 1} \!\! \mbox{1}}

\begin{document}

\title{Single-photon quantum nondemolition detectors constructed with
  \\ linear optics and projective measurements}   
\author{Pieter Kok\cite{pieter}, Hwang Lee, and Jonathan P.\ Dowling}
\address{Quantum Computing Technologies Group, Section 367 \\
  Jet Propulsion Laboratory, California Institute of Technology \\
  Mail Stop 126-347, 4800 Oak Grove Drive, Pasadena, California 91109}

\maketitle

\smallskip
\begin{abstract}
 Optical quantum nondemolition devices can provide essential tools
 for quantum information processing. Here, we describe several optical
 interferometers that signal the presence of a single photon in a
 particular input state without destroying it. We discuss both
 entanglement-assisted and non-entanglement-assisted interferometers,
 with ideal and realistic detectors. We found that existing
 detectors with 88\% quantum efficiency and single-photon resolution
 can yield output fidelities of up to 89\%, depending on the input
 state. Furthermore, we construct expanded protocols to perform QND
 detections of single photons that leave the polarization
 invariant. For detectors with 88\% efficiency we found
 polarization-preserving output fidelities of up to 98.5\%. 
 \medskip


\end{abstract}

\begin{multicols}{2}

%\section{Introduction}

Quantum Nondemolition (QND) devices measure observables and couple the
back-action noise only to the conjugate of the measured quantity. The
measurement projects the system under scrutiny onto an eigenstate of the
measured observable, and repeated QND measurements yield the same
outcome as the initial one. These devices therefore closely resemble
ideal von Neumann measurements \cite{vonneumann55}. QND devices can
be exploited, for example, to improve the sensitivity in 
gravitational wave detection, to create entanglement on demand, and
to implement QND-based quantum computing using projective measurements
(see Ref.\ \cite{grangier98,turchette95,maitre97,cerf98,knill01}). 

In quantum optics, we usually consider QND devices in the context of
photon-number measurements. There, we have the added complication of
ultimate demolition; common photodetectors destroy the
photons they detect. To avoid this demolition, there exist QND
proposals using special components such as Kerr media, parametric
amplifiers, cold atoms in magneto-optical traps, or cavity quantum
electrodynamics \cite{applications}. However, these protocols generally
require strong nonlinearities or are highly frequency-dependent. 

In a recent experiment, Nogues {\em et al}.\ performed a
single-photon QND measurement of a weak cavity field. They used the
resonant coupling between the field and atoms moving through
the cavity \cite{nogues99}. When there was a single photon in the
cavity, the initial atom-field state $|g,1\rangle$ evolved in time
according to $|g,1\rangle\rightarrow\cos(\Omega t/2)|g,1\rangle +
\sin(\Omega t/2)|e,0\rangle$. Here, $|g\rangle$ and $|e\rangle$ are the
ground state and the excited state of the atoms. The photon-number
field states are $|0\rangle$ and $|1\rangle$, and $\Omega$ is the Rabi
frequency. When $t = 2\pi/\Omega$, the initial state $|g,1\rangle$
aquired a phase shift of $\pi$. By contrast, when there were no
photons in the cavity, the state $|g,0\rangle$ would not accumulate a
phase shift. The relative phase-shift was then measured with a Ramsey
interferometer by coupling $|g\rangle$ with another atomic level
$|g'\rangle$ \cite{nogues99}.  

Since a Ramsey interferometer involves atomic transitions, this QND
device depends strongly on the frequency of the field. The next step
is therefore to investigate simple {\em frequency independent} QND
devices. One possibility is the use of non-deterministic
interferometric QND detection. 

This paper is organized as follows: in section \ref{sec:kerr} we
describe the conventional interferometric QND device that uses Kerr
nonlinearities, and we also present a simple teleportation-based scheme to
detect the presence of a single photon. In section \ref{sec:main} we
introduce an interferometric scheme based on linear optics and
projective measurements, and we study the effect of inefficient
detectors and degraded single-photon auxiliary input states. The
difference between this and the teleportation-based protocol is that
this scheme can detect a single photon out of 0, 1 or 2 photons,
whereas the teleportation-based scheme can tell the difference between
only 0 and 1 photon. This difference is critical for linear optical
quantum computing. Finally, in section \ref{sec:pol} we construct a
modified protocol to perform QND detections of single photons that
leave the polarization invariant. 

\section{Interferometric QND}\label{sec:kerr}

In this section we present the conventional interferometric quantum
nondemolition scheme that uses the Kerr effect \cite{imoto85}. This scheme
is based on a phase measurement, and the fundamental phase error puts
a bound on the strength of the Kerr nonlinearity. Secondly, we present
a simple scheme to achieve single-photon QND detection with linear
optics based on teleportation. 

\subsection{Kerr nonlinearities}

\begin{figure}[h]
  \begin{center}
  \begin{psfrags}
     \psfrag{d1}{$D_1$}
     \psfrag{d2}{$D_2$}
     \psfrag{a}{$a$}
     \psfrag{b}{$b$}
     \psfrag{Kerr medium}{Kerr medium}
     \psfrag{in}{$\!\!|\psi_{\rm in}\rangle$}
     \psfrag{pr}{}
     \epsfxsize=8in
     \epsfbox[-20 20 780 135]{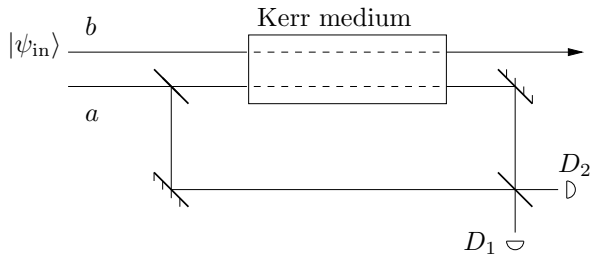}
  \end{psfrags}
  \end{center}
  \caption{A photon-number quantum nondemolition device based on the
     Kerr effect. A photon in mode $b$ changes the phase of the photon
     in mode $a$. The Mach-Zehnder interferometer is tuned in such a
     way that a detector click in $D_1$ signals no photon, and a
     click in $D_2$ signals one photon.} 
  \label{kerr}
\end{figure}

It is widely believed that to create an interferometric, 
photon-number quantum nondemolition device, you need a Kerr medium
\cite{imoto85}. A QND device based on such a nonlinearity works as
follows (see Fig.\ \ref{kerr}): let $\hat{a}^{\dagger}$,
$\hat{b}^{\dagger}$ and $\hat{a}$, 
$\hat{b}$ be the creation and annihilation operators for two optical
modes $a$ and $b$, satisfying the commutation relations
\begin{equation}
 [\hat{a},\hat{a}^{\dagger}] = [\hat{b},\hat{b}^{\dagger}] = 1
 \quad\text{and}\quad
 [\hat{a},\hat{a}] = [\hat{b},\hat{b}] = 0 \; . 
\end{equation}
The following Hamiltonian describes the effect of a Kerr cell on modes
$a$ and $b$: 
\begin{equation}
  \hat{H}_{\rm Kerr} = \kappa\, \hat{a}^{\dagger} \hat{a}\,\hat{b}^{\dagger}
  \hat{b} \; .
\end{equation}
This interaction induces a phase-shift in mode $a$ (or $b$), depending
on the number of photons in mode $b$ (or $a$). In general the mode
transformations are
\begin{equation}
  \hat{a}^{\dagger} \rightarrow \hat{a}^{\dagger}
  e^{-i\tau\hat{n}_b} \quad\text{and}\quad
  \hat{b}^{\dagger} \rightarrow \hat{b}^{\dagger}
  e^{-i\tau\hat{n}_a}\; ,
\end{equation}
where the dimensionless characteristic interaction strength $\tau
\equiv \kappa t/\hbar$ is based on the interaction time $t$ and the
number operator $\hat{n}_a$ (defined by $\hat{n}_a \equiv
\hat{a}^{\dagger}\hat{a}$ and similarly for $\hat{n}_b$). By
monitoring the phase shift in mode $a$ using a Mach-Zehnder
interferometer (i.e., homodyne detection), we can determine the
photon-number in mode $b$ {\em without} destroying the photons. The
phase of the output state is completely uncertain, which can be
understood from the number--phase uncertainty relation. For a detailed
description of general nonlinear interferometers, see e.g., the
article by Sanders and Rice \cite{sanders00}.    

We can measure any photon number in mode $b$ by
estimating the induced phase in mode $a$. However, the phase
estimation process has a fundamental error $\Delta\phi$ (where $\phi =
\tau\hat{n}_b$ is the induced phase). To attain
single-photon resolution, this error should be strictly smaller than
the phase shift induced by a single photon, or $\Delta\phi < \tau$. Of
course, we are not restricted to single-photon interferometry to
estimate this phase shift. We can use Heisenberg-limited
interferometry to obtain a lower bound on the strength of the Kerr
nonlinearity \cite{ou96}.  

For example, we may use so-called ``noon states" \cite{lee02}
$(|N,0\rangle + |0,N\rangle)/\sqrt{2}$, where $|0\rangle$ and $|N\rangle$
are the vacuum and the $N$-photon Fock states respectively. If the state
inside the Mach-Zehnder interferometer has this form, we obtain the
following bound for the phase uncertainty and the strength of the Kerr
effect:
\begin{equation}\label{error}
  \tau > \Delta\phi = \frac{\pi}{2} \frac{1}{N}\; .
\end{equation}
When we use single-photon interferometry ($N$=1), we retrieve the
well-known value of $\tau=\pi/2$. In experiments with coherent states,
the phase is determined by the standard limit, which yields $\tau >
\pi/(2\sqrt{\langle n\rangle})$. Here, $\langle n\rangle$ is the
average photon number. By using high-intensity laser beams, Grangier
{\em et al}.\ demonstrated Kerr-based QND measurements with small
nonlinearities \cite{grangier98}.    

Due to the typically small values of the nonlinearity, the
Kerr-based single-photon QND device is not practical. The $\chi^{(3)}$
coupling  involved is extremely weak ($\kappa\propto 10^{-16}$ cm$^2$ sV$^{-2}$
\cite{boyd99}), and such a detection device would necessarily have an
exceedingly small efficiency. Recently, large Kerr nonlinearities were
constructed using slow light, but these techniques are still highly
experimental \cite{lukin00}. We therefore wish to bypass the use of
weak $\chi^{(3)}$ nonlinearities and construct a single-photon QND
device with more 
user-friendly optical elements. In this paper, we show that under
certain relaxing conditions only linear optics and
projective measurements suffice.  

\subsection{Strength of the nonlinearity}

Before we continue, we derive the strength of the Kerr nonlinearity in
terms of the dimensionless coupling constant $\tau$ of
Eq.~(\ref{error}). A wave travelling through a medium accumulates a
phase shift $\varphi$ that in the scalar approximation is given by
$\varphi = \vec{k}\cdot\vec{x} = k L n$. The wave number is denoted by
$k$, $L$ is the length of the medium, and $n$ is the index of
refraction. In general, $n$ can be written in terms of the
higher-order nonlinearities as 
\begin{equation}
  n^2 = 1 + \chi^{(1)} + \chi^{(2)} E + \chi^{(3)} E^2 + \ldots,
\end{equation}
where $E$ is the electric field strength of the wave. Here, we are
interested in the third-order ($\chi^{(3)}$) contribution to the phase
shift, and we choose $\chi^{(1)} = \chi^{(2)} = 0$. The total phase
shift then becomes
\begin{equation}
  \varphi = k L \sqrt{1 + \chi^{(3)} E^2} \approx k L \left( 1 +
  \frac{\chi^{(3)} E^2}{2} \right)\; .
\end{equation}
The phase shift $\tau$ due to the Kerr effect is then 
\begin{equation}
  \tau = \frac{1}{2} k L \chi^{(3)} E^2\; .
\end{equation}

Since $\chi^{(3)}$ is a constant of the material, we need only 
to determine $E$ for a single photon to find the
numerical value of $\tau$. In the appropriate units the electric
field of a single photon with central frequency $\omega$ becomes
\cite{scully97} 
\begin{equation}
  E = \sqrt{\frac{\hbar\omega}{2 \epsilon_0 V}}\; ,
\end{equation}
where $\epsilon_0$ is the permittivity of the vacuum and $V$ the
volume of the medium that induces the phase shift. This yields 
\begin{equation}
  \tau = \frac{\hbar\omega^2 \Delta t \chi^{(3)}}{4 \epsilon_0 V}\; .
\end{equation}
Here, we used $k=\omega/c$ and $\Delta t = L/c$, where $c$ is the
speed of light in vacuum. When we choose the typical values of $\omega
= 3\cdot 10^{15}$~rad~s$^{-1}$, $\Delta t = 3\cdot 10^{-11}$~s,
$\chi^{(3)} = 2\cdot 10^{-22}$~m$^2$~V$^{-2}$, and the size of the
Kerr medium is 1~cm~$\times$~0.1~cm$^2$, the value for the
dimensionless coupling becomes $\tau \approx 10^{-18}$.

\subsection{Teleportation-based protocol}

\begin{figure}[h]
  \begin{center}
  \begin{psfrags}
     \psfrag{D1}{$D_1$}
     \psfrag{D2}{$D_2$}
     \psfrag{input}{$|\psi_{\rm in}\rangle = c_0
     |0\rangle + c_1 |1\rangle$}
     \psfrag{output}{$ |\psi_{\rm out}\rangle = |1\rangle$}
     \psfrag{PDC}{PDC}
     \epsfxsize=8in
     \epsfbox[-30 10 770 140]{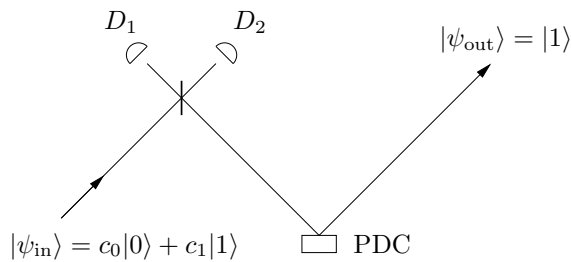}
  \end{psfrags}
  \end{center}
  \caption{Teleportation-based single-photon quantum nondemolition
     device. A detector coincidence in $D_1$ and $D_2$ signals the
     presence of a photon in $|\psi_{\rm in}\rangle$ and $|\psi_{\rm
     out}\rangle$. This scheme breaks down when there is potentially
     more than one photon in the input state.}
  \label{tel}
\end{figure}

One simple way to perform a single-photon measurement without
destroying the photon is to use single-photon quantum teleportation
\cite{bouwmeester97}. In Fig.\ \ref{tel} we show how such a protocol
would work. The input state $|\psi_{\rm in}\rangle$ may be in an
arbitrary superposition of zero and one photon with a particular
polarization. A maximally polarization-entangled photon-pair (e.g.,
created by a parametric down-converter, or PDC) serves as the quantum
channel, and a detector coincidence in $D_1$ and $D_2$ identifies a
(partial) Bell measurement. This detector coincidence also signals the
presence of a single photon in the input and the output state. It is
easily seen that a vacuum input state ($|\psi_{\rm in}\rangle =
|0\rangle$) can never lead to a two-fold detector coincidence, and
that a low efficiency pair production in the down-converter does not
affect the fidelity of the single-photon QND device. 

However, this scheme breaks down in the more interesting case where
the input state is of the form 
\begin{equation}\label{inputstate}
  |\psi_{\rm in}\rangle = c_0 |0\rangle + c_1 |1\rangle + c_2
   |2\rangle\; .
\end{equation}
The two-photon term will end up contributing to the coincidences in
$D_1$ and $D_2$ when the output of the down-converter is vacuum. This
scheme therefore falsely identifies the presence of a single photon
conditioned on a detector coincidence.

In the next section we shall develop an interferometer that faithfully
signals the presence of a single photon in such an input state. In
section \ref{sectel} we shall give a more detailed analysis of the
teleportation-based QND device, including unknown polarizations of the
input state. In the next section we study an interferometric scheme
that does work on an input state given by Eq.~(\ref{inputstate}).

\section{Linear optics and projective measurements}\label{sec:main}

Our main goal in this section is to describe an interferometric
single-photon QND device that non-destructively signals the presence
of a single photon when the input state is of the form of Eq.\
(\ref{inputstate}). Note that this is not a full QND measurement of
the photon-number observable since it works for only the lowest three
numbers (0, 1, and 2). However, it can still play an important r\^ole
in linear optical quantum computing, where up to only two-photon
states are used, such as the recently proposed linear optical quantum
computation scheme by Knill, Laflamme and Milburn \cite{knill01}. 

To construct such a dressed-down QND device, we consider the
following two relaxing conditions: Firstly, when the device does {\em
  not} signal the presence of a single photon, the output state may be
severely disturbed; we therefore essentially propose a device for
protocols using single-photon post-selection. Secondly, we do not
require a 100\% efficiency for the QND device. It is sufficient to
create a probabilistic device that has a lower effective efficiency.
Under these conditions, we show how to build a single-photon QND
device. 

The interferometric QND detector has possible applications in the
construction of quantum logic gates \cite{qgates}, single-photon triggers
\cite{trigger}, detectors of the quantum Zeno effect \cite{zeno},
quantum repeaters \cite{repeater} and to fundamental tests of quantum
mechanics \cite{fundamental}. In the following sections we shall
describe the interferometer by specifying the transformation
properties of the creation operators of the different optical modes,
and we calculate the efficiency and fidelity in the presence of
detector losses. Finally, we consider imperfections in the
single-photon probe sources.

\subsection{The interferometer}

\begin{figure}[h]
  \begin{center}
     \epsfxsize=8in
     \epsfbox[-80 20 970 230]{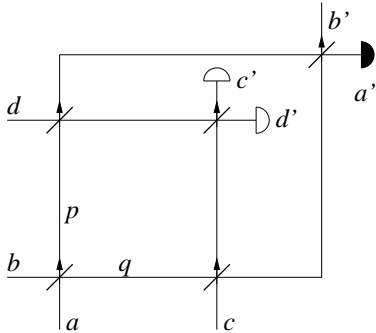}
  \end{center}
  \caption{Interferometric QND measurement device for single-photon
     detection. The device signals the presence of a single photon in
     mode $a$ by giving a detector coincidence in modes $c'$ and $d'$
     and {\em no} count in mode $a'$. The outgoing mode $b'$ then is
     in a single-photon state.}  
  \label{fig1}
\end{figure}

Consider the interferometric setup in Fig.\ \ref{fig1}. The beam
splitters are chosen asymmetric: When a photon is transmitted it
will not experience a phase shift, but when a photon is reflected it
accumulates a relative minus sign, depending on the side it reflects
off. The arrow in Fig.\ \ref{fig1} indicates the preferred direction: 
\begin{equation}\label{bs}
  \hat{a}^{\dagger} \rightarrow \frac{\hat{p}^{\dagger} +
  \hat{q}^{\dagger}}{\sqrt{2}} \quad\text{and}\quad \hat{b}^{\dagger}
  \rightarrow \frac{\hat{q}^{\dagger} - \hat{p}^{\dagger}}{\sqrt{2}}\; .
\end{equation}
You can convert these transformation rules into other representations
of the beam splitter by using three phase shifts. For example, a phase
shift of $\pi/2$ in modes $b$ and $q$, and a phase shift of $\pi$ in
mode $p$ yields 
\begin{equation}\nonumber
  \hat{a}^{\dagger} \rightarrow \frac{-\hat{p}^{\dagger} +
  i\hat{q}^{\dagger}}{\sqrt{2}} \quad\text{and}\quad \hat{b}^{\dagger}
  \rightarrow \frac{i\hat{p}^{\dagger} - \hat{q}^{\dagger}}{\sqrt{2}}\; .
\end{equation} 
In the rest of the paper we shall use Eq.~(\ref{bs}). The creation
operators $\hat{a}^{\dagger}$, $\hat{b}^{\dagger}$,
$\hat{c}^{\dagger}$ and $\hat{d}^{\dagger}$ of the input modes $a$,
$b$, $c$ and $d$ then transform into   
\begin{mathletters}
\begin{eqnarray}
  \hat{a}^{\dagger} & \rightarrow & \frac{\mbox{$\hat{a}'$}^{\dagger}
  - \mbox{$\hat{c}'$}^{\dagger}}{\sqrt{2}}\; , \label{a} \\
  \hat{b}^{\dagger} & \rightarrow & \frac{\mbox{$\hat{b}'$}^{\dagger}
  + \mbox{$\hat{c}'$}^{\dagger}}{\sqrt{2}}\; ,
  \label{b} \\
  \hat{c}^{\dagger} & \rightarrow & \frac{1}{2}\left(
  \mbox{$\hat{a}'$}^{\dagger} + \mbox{$\hat{b}'$}^{\dagger} +
  \mbox{$\hat{c}'$}^{\dagger} + \mbox{$\hat{d}'$}^{\dagger} \right)\;
  , \label{c} \\ 
  \hat{d}^{\dagger} & \rightarrow & \frac{1}{2}\left(
  \mbox{$\hat{b}'$}^{\dagger} - \mbox{$\hat{a}'$}^{\dagger} + 
  \mbox{$\hat{d}'$}^{\dagger} - \mbox{$\hat{c}'$}^{\dagger} \right)\;
  . \label{d} 
\end{eqnarray}
\end{mathletters}
We shall now study the behaviour of the QND device.

Let the input state be
\begin{equation}
  |\Psi_{\rm in}\rangle = |\psi_{\rm in},0,1,1\rangle_{abcd}\; ,
\end{equation}
where $|\psi_{\rm in}\rangle$ is defined by Eq.~(\ref{inputstate}). In
terms of creation operators, this state can be written as  
\begin{equation}\label{opinput}
  |\Psi_{\rm in}\rangle = \sum_{k=0}^2 c_k \hat{a}^{\dagger k}\;
   \hat{c}^{\dagger} \hat{d}^{\dagger}\; |0\rangle\; .
\end{equation}
Using Eqs.~(\ref{c}) and (\ref{d}), the two probe photons in modes $c$ and
$d$ transform into [see Eq.~(\ref{opinput})]
\begin{equation}\label{bog}
  \hat{c}^{\dagger}\hat{d}^{\dagger} \rightarrow \frac{1}{4}\left(
  \mbox{$\hat{b}'$}^{\dagger 2} - \mbox{$\hat{a}'$}^{\dagger 2} +
  \mbox{$\hat{d}'$}^{\dagger 2} - \mbox{$\hat{c}'$}^{\dagger 2} - 2
  \mbox{$\hat{a}'$}^{\dagger}\mbox{$\hat{c}'$}^{\dagger} + 2
  \mbox{$\hat{b}'$}^{\dagger} \mbox{$\hat{d}'$}^{\dagger} \right)\; . 
\end{equation}
Based on one photon in mode $c$, one in mode $d$, and nothing in $a$
and $b$, we can never have a coincidence in modes $c'$ and $d'$: there
is no $\hat{c}' \hat{d}'$ component in Eq.~(\ref{bog}). Similarly,
when the input mode is in a two-photon state $|2\rangle$ ($k=2$), the
operator transformation from Eq.~(\ref{a}) yields 
\begin{equation}\label{asquared}
  \hat{a}^{\dagger 2} \rightarrow \frac{1}{2} \left(
  \mbox{$\hat{a}'$}^{\dagger 2} - 2 \mbox{$\hat{a}'$}^{\dagger}
  \mbox{$\hat{c}'$}^{\dagger} + \mbox{$\hat{c}'$}^{\dagger 2}
  \right)\; .
\end{equation}
The only detector coincidence in modes $c'$ and $d'$ due to a
two-photon input is obtained when the $2\mbox{$\hat{b}'$}^{\dagger}
\mbox{$\hat{d}'$}^{\dagger}$ contribution from Eq.~(\ref{bog}) is
combined with the $2 \mbox{$\hat{a}'$}^{\dagger}
\mbox{$\hat{c}'$}^{\dagger}$ contribution of Eq.~(\ref{asquared}) to
yield $\mbox{$\hat{a}'$}^{\dagger} \mbox{$\hat{b}'$}^{\dagger}
\mbox{$\hat{c}'$}^{\dagger} \mbox{$\hat{d}'$}^{\dagger}$. However,
post-selecting on vacuum in mode $a'$ rules out this two-photon
contribution to the detector coincidence in $c'$ and $d'$.

Finally, a single photon in mode $a$ yields a contribution
$\mbox{$\hat{b}'$}^{\dagger} \mbox{$\hat{c}'$}^{\dagger}
\mbox{$\hat{d}'$}^{\dagger}$ [$k=1$ in Eq.~(\ref{opinput})]---that
is, there is a coincidence event in modes $c'$ and $d'$, {\em and}
there is a photon in the output mode $b'$. These properties constitute
our single-photon quantum nondemolition device. The efficiency of this
interferometric QND device with 50:50 beam splitters is 1/8. This can
be easily seen by determining the prefactor of the contribution
$\mbox{$\hat{b}'$}^{\dagger} \mbox{$\hat{c}'$}^{\dagger}
\mbox{$\hat{d}'$}^{\dagger}$, which yields $1/(2\sqrt{2})$. The square
of this amplitude is 1/8. We can optimize this probability by changing
the transmission coefficients $T$ of the beam splitters in modes $c$
and $d$. When we choose $T=1/3$, the probability of success becomes
4/27.  

\subsection{Realistic detectors}\label{sec:realdet}

For interferometers that operate at the low photon-number
level, the fidelity of the output state depends critically on the
performance of the photodetectors. The same is true for our
single-photon QND device. So far we have implicitly assumed that we
have ideal detectors---that is, the detectors give the correct photon
number in a mode every time they are used. Such detectors are said to
have unit efficiency and perfect single-photon resolution
\cite{kok01}. However, these detectors do not exist. It is therefore
important to study how imperfections in the detection process affect
the fidelity of the output state.

First, we define the fidelity of the output state of the
interferometer as  
\begin{equation}\label{fidelity}
  F \equiv {\mathrm Tr}[\rho_{\rm out} |\psi\rangle\langle\psi|] \leq 1\; ,
\end{equation}
where $|\psi\rangle$ is the expected pure state (in our case the
single-photon state $|1\rangle$, conditioned on the detector
signature) and $\rho_{\rm out}$ the (generally mixed) output
state. When $F=1$, the QND device works perfectly. We shall now
study how detector inefficiencies affect the fidelity.

Suppose that the detectors have a non-unit quantum efficiency
$\eta$, i.e., the probability that a photon is detected is
$\eta^2$. Furthermore, suppose that the detectors have perfect
single-photon resolution. It can then still mistake a two-photon state
for a single photon when one of the photons is not detected. Kim {\em
  et al}.\ developed such detectors, which operate at about 7~K with a
quantum efficiency of $\eta^2=88\%$ \cite{kim99}. We shall next
calculate the fidelity of the single-photon QND device using these
detectors.

Mathematically, we model detection as follows: The outgoing
state $\rho_{\rm out}$ is obtained by tracing over the (pure) multi-mode
output state just before detection $|\Psi\rangle$ and the
positive operator-valued measure (POVM) $\hat{E}_{\vec{n}}$:
\begin{equation}\label{rhoout}
  \rho_{\rm out} = {\mathrm Tr} \left[\hat{E}_{\vec{n}}|\Psi\rangle
  \langle\Psi| \right]\; ,
\end{equation}
where $\vec{n}=(n_1,\ldots,n_M)$ denotes the detector signature of
finding $n_k$ photons in mode $k$. The total number of detected modes
$M$ must be smaller than the total number of modes $N$ in the
interferometer. Our task is to find a general expression for 
$\hat{E}_{\vec{n}}$. 

Since, in our approximation, detectors operate on single modes, the POVM
$\hat{E}_{\vec{n}}$ will factor into $\hat{E}_{n_1} \otimes \cdots
\otimes \hat{E}_{n_M}$, where the separate POVM's $\hat{E}_{n_k}$ are the
measures for single-mode detectors. When the detectors are ideal, the
POVM's reduce to projection operators
\begin{equation}
  \hat{E}_{n_k}^{\rm (ideal)} = |n_k\rangle\langle n_k|\; .
\end{equation}
In general, the single-mode detector POVM has the form 
\begin{equation}
  \hat{E}_{n} = \sum_{k=n}^{\infty} d_{n,k}\, |k\rangle\langle k|\; ,
\end{equation}
where the sum runs from $n$, since in our model the detector cannot
detect more photons than there are present in the incoming beam. In other
words, we discard dark counts. Furthermore, for $\hat{E}_{n}$
to be a proper POVM, we require that 
\begin{equation}
  \sum_{n=0}^{\infty} \hat{E}_{n} = \unity\; ,
\end{equation}
where $\unity$ is the identity operator.

To find the expression for $\hat{E}_{n}$, we model the detector loss
by a beam splitter with transmission amplitude $\eta$. In this model
we assume that the detector detects photons independently. The
reflected photons represent the loss, and we take the trace over the
corresponding output mode. The POVM then becomes ($\tilde{\eta} \equiv
\sqrt{1-\eta^2}$) \cite{kok00}
\begin{eqnarray}\label{povm1}
  \hat{E}_{n} &=& {\mathrm Tr}_b \left[ \frac{1}{n!}
  (\eta\,\hat{a}^{\dagger} +  \tilde{\eta}\, \hat{b}^{\dagger})^n\,
  |0\rangle\langle 0|\, (\eta\,\hat{a} + \tilde{\eta}\, \hat{b})^n
  \right] \cr 
  &=& \sum_{k=0}^n
  \binom{n}{k}
  \; \eta^{2(n-k)}\tilde{\eta}^{2k}\;
  |n-k\rangle\langle n-k| \; .
\end{eqnarray}
Written like this, we have described the POVM for the transmission of
$n$ photons, which yields possible detector outcomes ranging from 0 to
$n$. However, we want the POVM corresponding to a {\em particular}
photon-number detector reading $k$. We therefore have to reverse the
r\^oles of $n$ and $k$ and adjust the summation in
Eq.~(\ref{povm1}). The POVM for detecting $k$ photons then becomes
\begin{equation}\label{povm2}
  \hat{E}_{k} = \sum_{n=k}^{\infty}\; \binom{n}{k}\; \eta^{2k}
  \tilde{\eta}^{2(n-k)}\; |n\rangle\langle n|\; .
\end{equation}
The lowest three POVM's are given by 
\begin{mathletters}
\begin{eqnarray}
  \hat{E}_0 &=& |0\rangle\langle 0| + \tilde{\eta}^2 |1\rangle\langle
  1| + \tilde{\eta}^4 |2\rangle\langle 2| + \ldots \\ && \cr
  \hat{E}_1 &=& \eta^2 |1\rangle\langle 1| + 2\eta^2\tilde{\eta}^2
  |2\rangle\langle 2| + 3\eta^2\tilde{\eta}^4 |3\rangle\langle 3| +
  \ldots \\ && \cr
  \hat{E}_2 &=& \eta^4 |2\rangle\langle 2| + 3\eta^4\tilde{\eta}^2
  |3\rangle\langle 3| + \ldots
\end{eqnarray}
\end{mathletters}
Alternatively, instead of calculating the coefficients $d_{k,l}$ of
$\hat{E}_k$ in a---necessarily simplified---model, these values can
in principle be determined experimentally. This would take into
account mure subtle effects, such as the saturation properties of the
detectors, and dark counts. Note that we can also include dark counts
in our model by inserting a thermal state into the secondary input
mode of the beam splitter. However, when we take detector readings
only during small time windows, the dark count contribution becomes
small, and the fidelity degradation is predominantly due to the
detector inefficiencies. 

\begin{figure}[h]
   \mbox{$~F$}\hfill
  \begin{center}
   \epsfxsize=8in
   \epsfbox[-30 60 820 210]{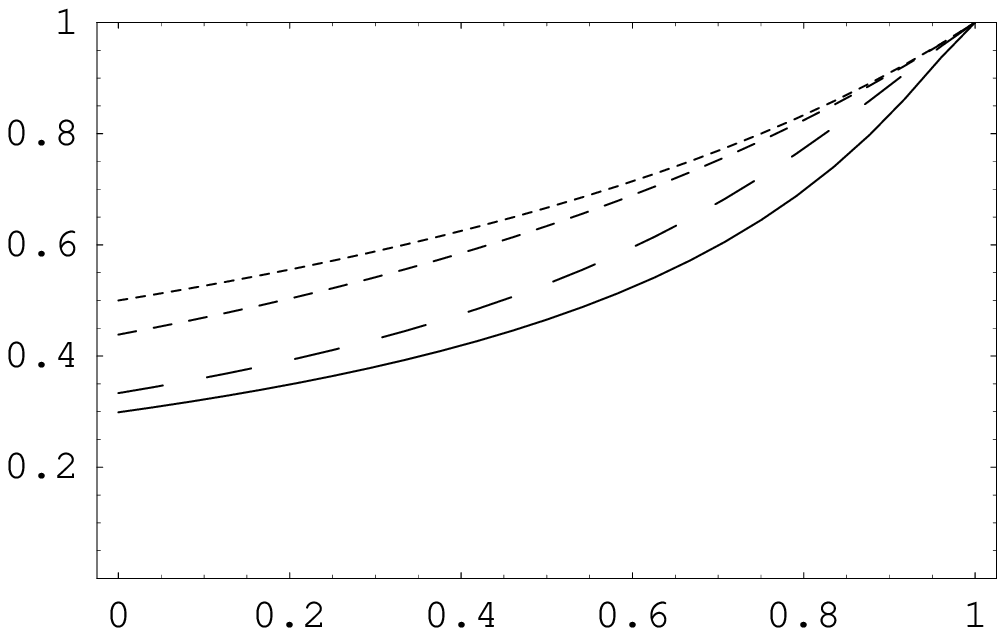}
   \mbox{$\eta$}
  \end{center}
  \caption{The fidelity $F$ as a function of the detector efficiency
   $\eta$ for different values of $\gamma$. From the upper (dotted)
   curve to the lower (continuous) curve, the values of $\gamma$ are
   0, 0.1, 1, and 10 respectively.}   
  \label{fig:fidelity}
\end{figure}

To find the fidelity of the single-photon QND device, we have
to combine Eq.~(\ref{povm2}) with Eqs.~(\ref{fidelity}) and
(\ref{rhoout}). For the output $\rho_{\rm out}$ in mode $b'$ this
yields 
\begin{equation}
  \rho_{\rm out} = {\mathrm Tr}_{a'c'd'}\left[\hat{E}_0^{(a')}\otimes
  \hat{E}_1^{(c')} \otimes \hat{E}_1^{(d')} |\Psi\rangle\langle\Psi|\right]\; ,
\end{equation}
that is, we condition the output on a detector coincidence in modes
$c'$ and $d'$, while nothing is detected in mode $a'$. When we
evaluate this expression we find the (unnormalized) output density
operator to be 
\begin{eqnarray}
  \rho_{\rm out} &=& \frac{1}{8} \left( \eta^4 |c_1|^2 + \frac{3}{2}\eta^4
  \tilde{\eta}^2|c_2|^2 \right) |1\rangle\langle 1| \cr
  && + \frac{1}{8} \left( \eta^4 \tilde{\eta}^2 |c_1|^2 + 3 \eta^4
  \tilde{\eta}^2 \right) |c_2|^2 |0\rangle\langle 0|\; .
\end{eqnarray}
The normalization factor is given by 
\begin{equation}\nonumber
  \frac{\eta^4}{8} \left[ (1+\tilde{\eta}^2) |c_1|^2 + \left(
  \frac{5}{2} \tilde{\eta}^2 + 6\tilde{\eta}^4 \right) |c_2|^2 \right] \; ,
\end{equation}
and the fidelity becomes
\begin{equation}\label{eq:fidelity}
  F = \frac{2 + 5\tilde{\eta}^2 \gamma}{2 + \tilde{\eta}^2 (2 + 5\gamma
  + 12\gamma\tilde{\eta}^2)}\; ,
\end{equation}
where $\gamma\equiv|c_2|^2/|c_1|^2$ is the two-photon fraction in the
input state with respect to the single-photon contribution. This
fidelity depends on the input state, characterized by $\gamma$. For
ideal detectors ($\eta^2=1$) we find $F=1$. In Fig.~\ref{fig:fidelity}
we plotted the fidelity of the single-photon QND device as a function
of the detector efficiencies for different values of $\gamma$. When we
use the detectors from Ref.~\cite{kim99} with $\eta^2 = 88\%$, the
fidelity is 89\% for $\gamma=0$ and $\gamma=0.1$, 86\% for
$\gamma=1$, and 80\% for $\gamma=10$. The output state of the
single-photon QND device is then 
\begin{equation}
  \rho_{\rm out} = (1-F)|0\rangle\langle 0| + F |1\rangle\langle 1|\; .  
\end{equation}

It is immediately clear from Fig.~\ref{fig:fidelity} that this
protocol works only when the detectors have a high quantum
efficency. In particular, when there is a sizeable two-photon
contribution in the input state, the fidelity remains below 50\% for
most of the efficiency domain. 

\subsection{Single-photon probes}

We based the above fidelity calculation on perfect single-photon sources
for the probe modes. When these modes are weak coherent states,
the output fidelity deteriorates considerably. This is because we use
{\em two} probes: For weak coherent states these
two single-photon inputs occur with approximately the same probability as
events with two probe photons in one mode and vacuum in the
other. However, when we employ parametric down-conversion to create
the two-photon states that act as probes, this problem disappears. 
Given that the parametric down-converter creates the output state
$(1-\epsilon^2)|0\rangle + \epsilon|1,1\rangle + O(\epsilon^2)$, the
vacuum contribution of this state is eliminated in the post-selection
process. Also, the efficiency of the whole QND device deteriorates
rapidly (to order $\sim 10^{-4}$), but this is still twelve orders of
magnitude better than using bulk Kerr nonlinearities. 

\section{Polarization preserving QND}\label{sec:pol}

So far, we have considered only optical fields in certain
photon-number superpositions. You can argue that we can construct a
single-photon ``semi-QND'' scheme by detecting the mode with a
single-photon resolution detector and subsequently creating another
photon with a single-photon gun. Indeed, this would work perfectly. 
However, when the incoming field has a certain unknown polarization
that needs to be preserved, the semi-QND scheme breaks down. Our next
goal is therefore to create an interferometric single-photon QND
device that preserves the polarization of a photon $|\theta\rangle
\equiv \alpha |H\rangle + \beta |V\rangle$, where $\alpha$ and $\beta$
are two complex numbers and $H,V$ the polarization directions. 

\subsection{Teleportation-based protocol}\label{sectel}

\begin{figure}[h]
  \begin{center}
  \begin{psfrags}
     \psfrag{D1}{$D_1$}
     \psfrag{D2}{$D_2$}
     \psfrag{input}{$|\psi_{\rm in}\rangle = c_0
     |0\rangle + c_1 |\theta\rangle$}
     \psfrag{output}{$ |\psi_{\rm out}\rangle = |\theta\rangle$}
     \psfrag{PDC}{PDC ($\chi^{(2)}$)}
     \epsfxsize=8in
     \epsfbox[-30 15 770 140]{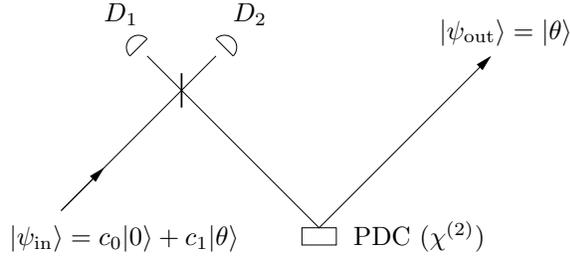}
  \end{psfrags}
  \end{center}
  \caption{Teleportation-based QND measurement device for
     single-photon detection that leaves the polarization of the
     photon intact. The device signals the presence of a single photon
     $|\theta\rangle$ in the input mode with (three-dimensional)
     polarization angle $\theta$, by giving a detector coincidence in
     $D_1$ and $D_2$. }
  \label{fig2}
\end{figure}

A simple and elegant way to create a polarization preserving QND
device is to {\em teleport} the polarization state with the protocol
used by Bouwmeester {\em et al.}\ \cite{bouwmeester97}. As shown in
Fig.\ \ref{fig2}, the  incoming state $|\psi_{\rm in}\rangle = c_0
|0\rangle + c_1 |\theta\rangle$ (where $|\theta\rangle =
\alpha|H\rangle + \beta|V\rangle$) is mixed in a beam splitter with
one half of a polarization entangled state from a parametric
down-converter, $|\Psi_{\rm   PDC}\rangle = (1-\epsilon^2)|0\rangle +
\epsilon (|H,V\rangle - |V,H\rangle)/\sqrt{2} + O(\epsilon^2)$. 
Post-selection on a two-fold coincidence (the partial Bell detection) in
detectors $D_1$ and $D_2$ yields the outgoing state $|\theta\rangle$.

The detector coincidence identifies the singlet state $|\Psi^-\rangle
\equiv (|H,V\rangle - |V,H\rangle)/\sqrt{2}$. The complete set of Bell
states is given by  
\begin{eqnarray}
  |\Psi^{\pm}\rangle &=& (|H,V\rangle \pm |V,H\rangle)/\sqrt{2}\; ,
   \cr
  |\Phi^{\pm}\rangle &=& (|H,H\rangle \pm |V,V\rangle)/\sqrt{2}\; .
\end{eqnarray}
It is well known that a deterministic complete Bell state detection is
impossible using linear optics \cite{lutkenhaus99}, but we can
construct a {\em  probabilistic} Bell measurement that works with
probability 1/2 using Ref.\ \cite{braunstein}. Unfortunately, the
teleportation protocol breaks down when there is a (sizeable) two-photon
contribution in the input state: When the incoming state is 
$|\psi_{\rm in}\rangle = c_0 |0\rangle + c_1|\theta\rangle + c_2
|2\theta\rangle$ (where $|2\theta\rangle$ is the two-photon Fock state
in the polarization mode $\theta$), then the outgoing state $\rho_{\rm
  tb}$ based on a two-fold detector-coincidence (to first order in
$p_{\rm pdc}$) is
\begin{eqnarray}
  \rho_{\rm tb} &=& \frac{3 p_{\rm pdc}\, |c_1|^2}{4|c_2|^2 + 3 p_{\rm
      pdc}\, |c_1|^2}\; |\theta\rangle\langle\theta| \cr
 && \qquad\qquad + \frac{4|c_2|^2}{4|c_2|^2 + 3 p_{\rm pdc}\, |c_1|^2}\,
  |0\rangle\langle 0| \; ,
\end{eqnarray}
where $p_{\rm pdc}\sim 10^{-4}$ is the probability of creating a single
polarization-entangled photon pair in the parametric down-converter. 
The fidelity $F={\rm Tr}[\rho_{\rm tb}|\theta\rangle\langle\theta|]$
becomes vanishingly small when the two-photon contribution in the
state increases (i.e., when $|c_2|^2 \gg p_{\rm pdc}\, |c_1|^2$). In
the next section we shall study an interferometer that signals the
presence of a single photon with arbitrary polarization even when
there is a sizeable two-photon contribution present in the input
state. 

\subsection{Two-photon robustness}

We have just demonstrated that the teleportation-based protocol to perform
polarization-invariant single-photon QND detections breaks down in the
presence of a sizeable two-photon amplitude. Analogous to section
\ref{sec:main} we now construct an interferometer that signals
the presence of a single photon with arbitrary polarization when the
input state is
\begin{equation}\label{polinputstate}
  |\psi_{\rm in}(\theta)\rangle = c_0 |0\rangle + c_1 |\theta\rangle +
   c_2 |2\theta\rangle\; , 
\end{equation}
where $|\theta\rangle$ is a single photon with a polarization angle
$\theta$. Note that $\theta$ might be a two-dimensional vector so as
to span the complete Bloch sphere. The term $|2\theta\rangle$ denotes
a two-photon state in the polarization mode corresponding to $\theta$.

\begin{figure}[h]
  \begin{center}
     \epsfxsize=8in
     \epsfbox[-50 10 750 220]{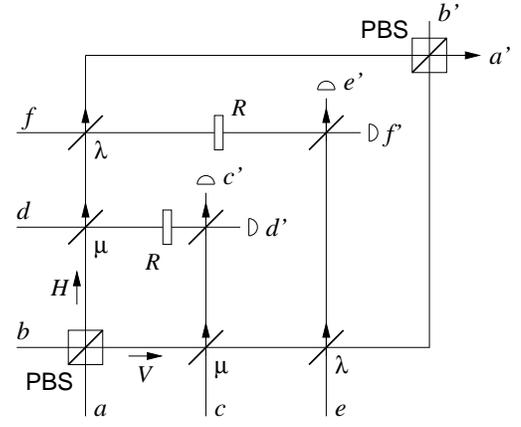}
  \end{center}
  \caption{The QND device for single-photon
     detection that leaves the polarization of the photon
     intact. Here, PBS denotes a polarization beam splitter and 
     $R$ is a polarization rotation over $\pi/2$ to erase the which-path
     information in the detectors. A four-fold detector coincidence in
     modes $c'$, $d'$, $e'$, and $f'$ (with one and only one photon
     per mode) signals the presence of a single photon in the input
     mode, while conserving the (unknown) polarization. The beam
     splitter coefficients are given by $\mu = \lambda = \frac{1}{2}
     \arccos(-1/3)$. The beam splitters before modes $c'$, $d'$, and
     $e'$, $f'$ have transmision coefficients 1/2.} 
  \label{fig4}
\end{figure}

Since the input mode can be in an arbitrary polarization state, we
describe this as a two-mode input. The interferometer in
Fig.~\ref{fig1} then works only if the single-photon auxiliary
input states in modes $c$ and $d$ have the same polarization as mode
$a$. As a consequence, this protocol cannot be used when the input
polarization is unknown. Consider therefore the interferometric setup
in Fig.~\ref{fig4}.

Since the beam splitter can be represented by a simple $SU(2)$
rotation around a fixed axis, we can parametrize the transmission
coefficient $T$ by the rotation angle $\mu$ such that
$T=\cos^2\mu$. When we choose the angles $\mu$ and $\lambda$ for the
different beam splitters according to  
\begin{equation}
  \mu = \lambda = \frac{1}{2} \arccos\left( -\frac{1}{3} \right)
  \quad\text{or}\quad T = \frac{1}{3} \; ,
\end{equation}
the mode transformations of this interferometer are
\begin{mathletters}
  \begin{eqnarray}
   \hat{a}^{\dagger}_H & \rightarrow & \frac{1}{3}\left[
   \mbox{$\hat{a}'$}^{\dagger}_H-\sqrt{3}\,(\mbox{$\hat{c}'$}^{\dagger}
   - \mbox{$\hat{d}'$}^{\dagger}) - \mbox{$\hat{e}'$}^{\dagger} +
   \mbox{$\hat{f}'$}^{\dagger} \right]\; , \label{pah} \\ 
   \hat{a}^{\dagger}_V & \rightarrow & \frac{1}{3}\left[
   \mbox{$\hat{a}'$}^{\dagger}_V-\sqrt{3}\,(\mbox{$\hat{c}'$}^{\dagger}
   + \mbox{$\hat{d}'$}^{\dagger}) - \mbox{$\hat{e}'$}^{\dagger} -
   \mbox{$\hat{f}'$}^{\dagger} \right]\; ,
   \label{pav} \\
   \hat{c}^{\dagger}_V & \rightarrow & \frac{
   2 \mbox{$\hat{a}'$}^{\dagger}_V + \sqrt{3}\,(\mbox{$\hat{c}'$}^{\dagger} + 
   \mbox{$\hat{d}'$}^{\dagger}) -2 (\mbox{$\hat{e}'$}^{\dagger} +
   \mbox{$\hat{f}'$}^{\dagger})}{3\sqrt{2}}\; , \label{pc} \\  
   \hat{d}^{\dagger}_H & \rightarrow & \frac{ - 2
   \mbox{$\hat{a}'$}^{\dagger}_H - \sqrt{3}\,
   (\mbox{$\hat{c}'$}^{\dagger} - \mbox{$\hat{d}'$}^{\dagger}) +2
   (\mbox{$\hat{e}'$}^{\dagger} -
   \mbox{$\hat{f}'$}^{\dagger})}{3\sqrt{2}}\; , \label{pd} \\ 
   \hat{e}^{\dagger}_V & \rightarrow & \frac{ 2
   \mbox{$\hat{a}'$}^{\dagger}_V + \mbox{$\hat{e}'$}^{\dagger} +
   \mbox{$\hat{f}'$}^{\dagger}}{\sqrt{6}}\; , \label{pe} \\ 
   \hat{f}^{\dagger}_H & \rightarrow & \frac{ - 2
   \mbox{$\hat{a}'$}^{\dagger}_H - \mbox{$\hat{e}'$}^{\dagger} + 
   \mbox{$\hat{f}'$}^{\dagger}}{\sqrt{6}}\; . \label{pf} 
  \end{eqnarray}
\end{mathletters}
In this set of equations we dropped the index $H,V$ of modes $c'$,
$d'$, $e'$, and $f'$, since the polarizations of these modes are
always identical.  

Let the input state in mode $a$ be given by Eq.~(\ref{polinputstate}), 
and feed probe photons with specified polarizations into modes $c$,
$d$, $e$, and $f$. The total input state $|\Psi_{\rm in}\rangle$ is
then given by 
\begin{eqnarray}
  |\Psi_{\rm in}\rangle &=& |\psi_{\rm in}(\theta),0,V,H,V,H
   \rangle_{abcdef} \cr &=& \sum_{k=0}^2 c_k\, \hat{a}^{\dagger
   k}_{\theta}\, \hat{c}_V^{\dagger} \hat{d}_H^{\dagger}
   \hat{e}_V^{\dagger} \hat{f}_H^{\dagger}\, |0\rangle_{abcdef}\; .
\end{eqnarray}
It is now a somewhat lengthy (but straightforward) calculation to show
that when ideal detectors in modes $c'$, $d'$, $e'$, and $f'$ all
record a single photon, the input state collapses onto the single-photon
state $|\theta\rangle\langle\theta|$ in the output mode $a'$. Note
that, due to the arrangement of the polarization beam splitters (PBS),
the detection of mode $b'$ is unnecessary: We do not have the
problematic notion of conditioning on non-detection. The probability
of a four-fold detector coincidence for a single-photon input in this
polarization preserving single-photon QND device is $(4/27)^2 \approx
2\%$. 

\begin{figure}[h]
   \mbox{$~F_{\theta}$}\hfill
  \begin{center}
   \epsfxsize=8in
   \epsfbox[-30 60 820 210]{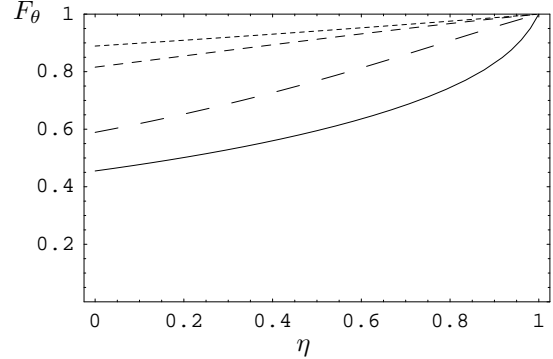}
   \mbox{$\eta$}
  \end{center}
  \caption{The fidelity $F_{\theta}$ as a function of the detector efficiency
   $\eta$ for different values of $\gamma$. From the upper (dotted)
   curve to the lower (continuous) curve, the values of $\gamma$ are
   0, 0.1, 1, and 10 respectively.}   
  \label{fig:polfidelity}
\end{figure}

Suppose that the input field is in a polarization mode with
$|\theta\rangle = (|H\rangle + |V\rangle)/\sqrt{2}$. Furthermore,
define $|\theta_{\perp}\rangle \equiv (|H\rangle -
|V\rangle)/\sqrt{2}$. When we use the realistic detectors modelled in
Sec.~\ref{sec:realdet} we find the outgoing state to be approximately
\begin{eqnarray}
  \rho_{\rm out} &\propto& \left( \frac{\tilde{\eta}^2}{365} + \gamma\frac{
  \tilde{\eta}^4}{49} \right) |0\rangle\langle 0| \cr 
  && ~ + \left(\frac{16}{729} + \gamma \frac{\tilde{\eta}^2}{49} \right) 
  |\theta\rangle\langle\theta| + \gamma \frac{\tilde{\eta}^2}{156}
  |\theta_{\perp}\rangle\langle \theta_{\perp}|\; ,
\end{eqnarray}
where $\gamma \equiv |c_2|^2/|c_1|^2$. For brevity, we have approximated the
lengthy algebraic terms in this expression by fractions that are
within 1\% of their numerical value. Furthermore, the appearance of
the term $|\theta_{\perp}\rangle\langle\theta_{\perp}|$ does not
indicate a polarization rotation associated with this interferometer;
it merely indicates that imperfect detections tens to randomize the
polarization of the output. After normalization the fidelity becomes
approximately 
\begin{equation}
  F_{\theta} \approx \frac{1 + 0.93\, \gamma\tilde{\eta}^2}{1 + (0.12 + 1.22\,
  \gamma)\tilde{\eta}^2 + 0.93\,\gamma \tilde{\eta}^4}\; .
\end{equation}
We have plotted $F_{\theta}$ as a function of the detector efficiency
is shown in Fig.~\ref{fig:polfidelity} for different values of
$\gamma$. A property we immediately notice about this interferometer
is that the fidelity is significantly more resilient to detector
losses than the single-photon QND device for fixed polarization (see
Fig.~\ref{fig:fidelity}). When we calculate the fidelities of the
output using the four different values of $\gamma$, we found that with
$\eta^2 = 88\%$, the fidelity is 98.5\% for $\gamma=0$, 98\% for
$\gamma=0.1$, 95\% for $\gamma=1$, and 81\% for $\gamma=10$. 

We now have two criteria to select between teleportation based
and interferometric QND detection: When we need a high-fidelity
single-photon QND detection and the two-photon contribution is negligible,
then we should use the teleportation-based quantum nondemolition
device (Fig.~\ref{fig2}). However, when we are in possession of
single-photon resolution detectors, or we need to exclude a
sizeable two-photon contribution, the interferometric methods given in
Figs.~\ref{fig1} and \ref{fig4} are superior. 

\section{Conclusions}

We have presented four, single-photon, quantum-nondemolition devices;
two were based on polarization entanglement as an auxiliary input and
two rely on single photons as auxiliary input states. A simple
teleportation-based protocol allows us to perform a single-photon QND
detection when the two-photon contribution in the input state is
negligible. We presented an interferometer based on single-photon
auxiliary input states and a two-fold detector coincidence that still
works when the main input mode is populated by two photons. The
optimal efficiency of this protocol is 4/27 or approximately 15\%. In
addition, we studied the effect of realistic detectors on the fidelity
of the interferometer, and we found that existing detectors with 88\%
quantum efficiency and single-photon resolution may yield a fidelity
of up to 89\%, depending on the input state. 

In the more general case where the polarization of the input mode
matters, we can again use teleportation as our single-photon QND
device. However, this protocol is still sensitive to two-photon
pollution. We therefore constructed an interferometer that is both
polarization preserving and that is robust against two-photon input
states. This setup involves four-fold coincidence detection, and the
efficiency is approximately 2\%. However, this interferometer is not
dependent on non-detection, contrary to the previous one, and as a
consequence, the fidelity of the output state is considerably higher. For
detectors with 88\% efficiency we found output fidelities of up to
98\%.  

All of the above protocols are inherently frequency-independent. The
question is now whether we can modify these schemes such that
two-photon inputs and general $n$-photon states can be identified
without destroying them. This is the object of further study.  

\section*{Acknowledgements}

This work was carried out at the Jet Propulsion Laboratory, California
Institute of Technology, under a contract with the National
Aeronautics and Space Administration. In addition, P.K.\ and H.L.\
acknowledge the United States National Research Council. Support was
received from the Advanced Research and Development Activity, the
National Security Agency, the Defense Advanced Research Projects
Agency, and the Office of Naval Research.

\end{multicols}
\end{document}